\documentclass[runningheads]{llncs}
\usepackage[T1]{fontenc}
\usepackage{comment}
\usepackage{graphicx}
\usepackage{amsmath}
\usepackage{amssymb}
\usepackage{hyperref}
\usepackage{booktabs} 
\usepackage{multirow} 
\begin{document}
\title{The Affective Bridge: Preserving Speech Representations while Enhancing Deepfake Detection via emotional Constraints}
%
%
%

\author{Yupei Li\inst{1,3}\thanks{Corresponding author: \email{yl7622@ic.ac.uk}} \and
Chenyang Lyv\inst{2} \and
Longyue Wang\inst{2} \and
Weihua Luo\inst{2} \and
Kaifu Zhang\inst{2} \and
Bj\"orn W. Schuller\inst{1,3}}

\authorrunning{Y. Li et al.}

\institute{Imperial College London, London, UK\\
\email{yl7622@ic.ac.uk}
\and
Alibaba International Digital Commerce Group\\
\email{\{lyuchenyang.lcy,wanglongyue.wly,weihua.luowh,kaifu.zhangkf\}@alibaba-inc.com}
\and
Technical University of Munich, Munich, Germany\\
\email{schuller@tum.de}}
\maketitle              
\begin{abstract}
Speech deepfake detection (DFD) has benefited from diverse acoustic and semantic speech representations, many of which encode valuable speech information and are costly to train. Prior work has shown that affective cues improve DFD, yet existing approaches either fuse emotion with other task-specific features in complex pipelines or directly fine-tune representations toward DFD objectives, risking distortion of the original speech representations that support downstream tasks such as speaker verification (SV) or automatic speech recognition (ASR). We propose a simpler approach: fine-tuning speech encoders on emotion recognition alone—without any DFD supervision, and training a lightweight support vector machine (SVM) on the frozen emotion-tuned representations for DFD. This preserves the original representation capacity for downstream tasks such as SV and ASR, while emergently improving DFD performance. Crucially, we find that emotion is uniquely effective as this bridging task: replacing it with speaker identity even degrades DFD performance, demonstrating that the benefit stems from emotion's role as a natural bridge between speech representation and DFD. Experiments on FakeOrReal and In-the-Wild show accuracy improvements of up to 6\% and 2\% with corresponding EER reductions, while analysis on ASVspoof 2019 LA reveals dataset-specific speaker bias in the real-speech subset. Code is available at \href{https://github.com/myxp-lyp/The-Affective-Bridge-Unifying-Feature-Representations-for-Speech-Deepfake-Detection}{GitHub repository}.

\end{abstract}
\section{Introduction}

Speech deepfake detection (DFD) in audio heavily relies on the selection of appropriate and robust feature representations.  Physical acoustic features are among the most commonly used, including Mel Frequency Cepstral Coefficients (MFCCs)~\cite{hamza2022deepfake}, chroma features~\cite{bakken2025deep}, and comprehensive sets of acoustic descriptors extracted using tools such as openSMILE~\cite{eyben2010opensmile}. In parallel, deep learning–based features derived from pretrained models such as Whisper~\cite{radford2023robust}, regarded as raw audio physical features, have also shown promising results. Beyond these low-level representations, high-level task-oriented features have proven effective as well. These are often obtained from models trained for specific downstream applications, such as automatic speech recognition (ASR)~\cite{mansoor2024audio} and speaker verification (SV)~\cite{pianese2022deepfake}, and tend to offer more representations with semantic information compared to simple physical features.

Given the extracted features, a common approach is to train the entire network end-to-end by jointly optimising the feature extractors and the downstream classifier. Under this setting, pretrained models such as Whisper or wav2vec 2.0 \cite{luo2024whisper,wang2025awaveformer} are directly fine-tuned for DFD, resulting in substantial structural adaptation of the original representations toward DFD features. While effective for detection, full fine-tuning substantially alters the feature space and limits the reuse of these representations in future tasks. Preserving their semantic information is therefore crucial, as the same features may later be required after DFD for tasks such as ASR and SV in multimodal large models \cite{xu2025qwen2}. Moreover, end-to-end fine-tuning typically needs to be repeated for each new deepfake dataset, as the learned representations may overfit to dataset-specific characteristics \cite{khanjani2023audio}. On the other hand, to reduce training cost, an alternative post-hoc strategy directly applies a classifier head on top of frozen speech representations for DFD \cite{saha2024exploring}. While this approach is computationally efficient, it treats the extracted features as fixed and provides no structured mechanism to guide the representations toward deepfake-discriminative characteristics.

Despite these advances, the field of audio DFD remains fragmented, with no unified training framework that both preserves the semantic information of the original features and guides them toward deepfake-discriminative representations. Notably, we observe that the emotion recognition task provides a shared factor that can guide heterogeneous speech representations toward deepfake discrimination. Firstly, emotion can be expressed through a variety of feature sets, including low-level descriptors of physical characteristics~\cite{li2025gatedxlstmmultimodalaffectivecomputing} as well as higher-level application-based features derived from ASR systems~\cite{tits2018asr}. For instance, the Macro-Voice framework disentangles speaker identity from emotional cues to obtain a more purified emotion representation, highlighting the intersection between speaker verification representations and emotional features~\cite{tian2025marcovoicetechnicalreport}. These observations suggest that most acoustic features capture emotional information and can be used for emotion recognition. Moreover, emotion cues have been shown to provide effective discriminative information for audio DFD~\cite{conti2022deepfake,mittal2020emotions} and are inherently difficult for generative models to reproduce authentically over successive iterations, making them a robust and reliable signal for DFD in practical applications~\cite{sun2024friendlyaicomprehensivereview,li2025artificialemotionsurveytheories}. Therefore, emotion recognition can act as a bridge for leveraging heterogeneous speech features in DFD.

To address the lack of a unified, feature-preserving approach for audio DFD, we make the following \textbf{contribution}. We propose a \textbf{training pipeline that introduces a pre-training stage prior to the final post-hoc classifier}, without fine-tuning the feature extractors and requiring only a lightweight classifier for detection.  In the pre-training stage, we found \textbf{emotion}, rather than other tasks such as identity recognition, as a suitable feature-agnostic constraint applicable to arbitrary speech representations, \textbf{guiding them toward deepfake-discriminative cues while preserving their original semantic structure, making it a safe and lightweight supplement to existing speech representations rather than a replacement}. This design enables deepfake-oriented representation adjustment with lower training cost than full fine-tuning.

Our method \textbf{differs fundamentally} from prior work that leverages emotion for DFD through \textbf{multimodal alignment}~\cite{hosler2021deepfakes}, where visual or other modalities are incorporated to enrich emotion representations that are then directly used as detection features. Our method also differs from \textbf{explicit fusion} strategies that concatenate or select emotion features alongside other audio representations~\cite{alsaeedi2025audio}. Such approaches require careful feature engineering and scale poorly as feature diversity grows. In contrast, we treat emotion not as a feature to be fused, but as a supervisory signal for fine-tuning: by adapting speech encoders through emotion recognition alone, without any DFD supervision, we allow emotion-relevant signal to emerge within the representation itself. Unlike approaches that further fine-tune the encoder on DFD objectives after emotion adaptation, we impose no DFD-specific supervision on the representation at any stage. Instead, a simple support vector machine (SVM) trained on the frozen emotion-tuned features suffices for detection, requiring minimal computational overhead, avoiding any distortion of the learned representation, and preserving the encoder's capacity for downstream tasks such as SV and ASR. Experimental results show consistent improvements on the FakeOrReal (FoR)~\cite{reimao2019dataset} and InTheWild (ITW)~\cite{muller2022does} benchmarks, specifically up to approximately 6\% and 2\% increases for accuracy, respectively, and in equal error rate (EER), showing reductions of up to about 4\% and 1\%, and comparable performance on ASVSpoof2019 LA~\cite{wang2020asvspoof}.


\section{Methodology: Affective Bridge for deepfake detection}
\subsection{Overview}
\begin{figure*}[ht]
    \centering
    \includegraphics[width=0.8\linewidth]{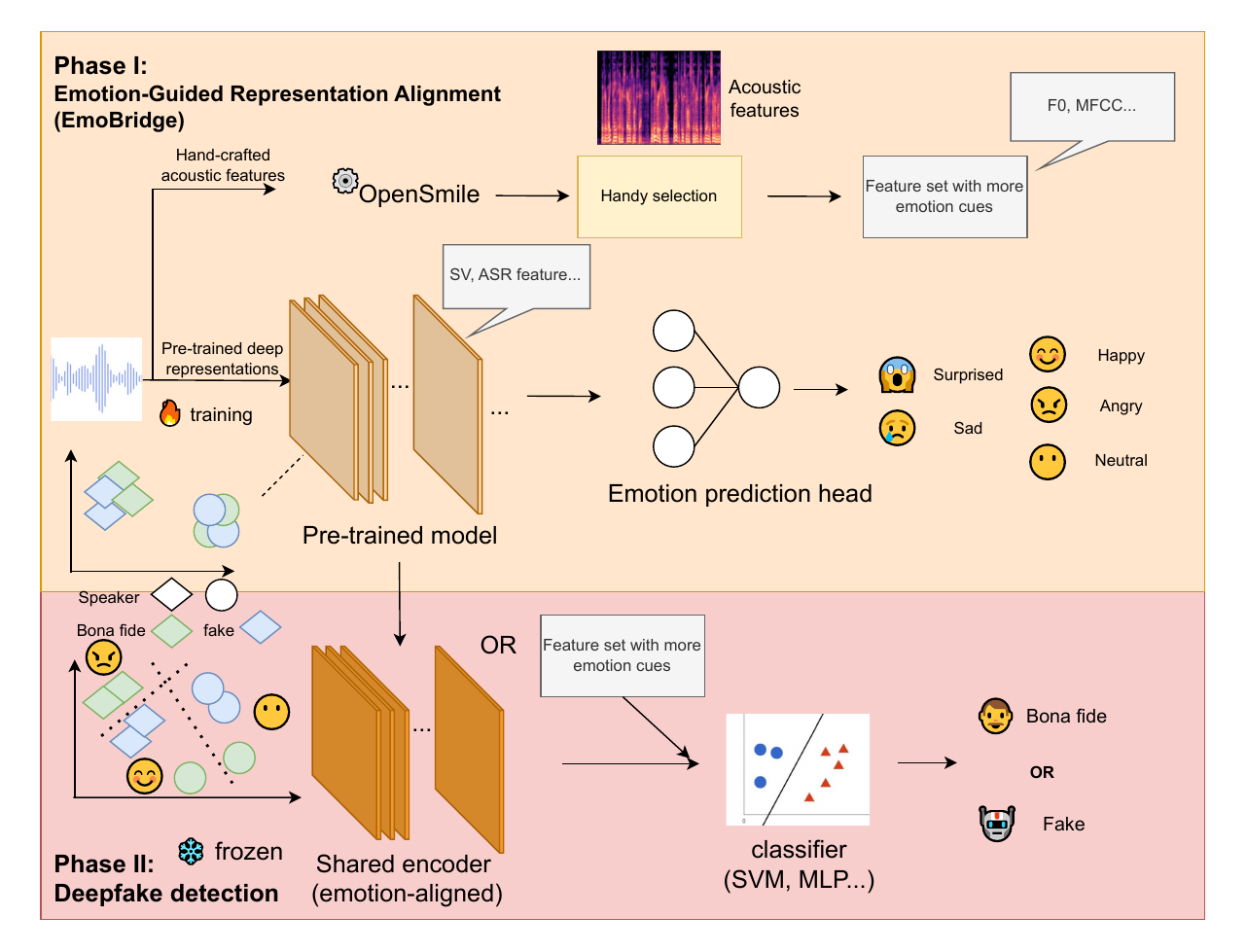}
    \caption{Overview of the proposed Affective Bridge framework for speech DFD. In Phase I Emotion-Guided Representation Alignment (EmoBridge), raw speech is transformed into speech representations through two parallel routes: hand-crafted acoustic feature extractors (e.g., OpenSMILE) or encoders pre-trained for other speech tasks such as ASR or speaker verification. For hand-crafted acoustic features, a subset of features containing richer emotion-related cues is retained. For deep representations, an emotion prediction head is attached to the pre-trained encoder and optimized using an emotion recognition objective, guiding the representation to incorporate emotion-related information while preserving its original task-oriented structure. In Phase II deepfake detection, the resulting emotion-aligned representation, either selected acoustic features or outputs from the trained encoder, are frozen and fed into a downstream classifier (e.g., SVM or MLP) to perform DFD, classifying input speech as real or fake.}
    \label{fig:framework}
\end{figure*}

To address the aforementioned challenge of unifying a training pipeline for preserving speech representation to enhance DFD, we propose Affective Bridge for deepfake detection, a two-phase framework that enhances speech DFD without any DFD supervision during representation learning. Our key insight is that emotion recognition serves as a natural intermediate task: by fine-tuning speech encoders on emotion alone, their representations emergently become more discriminative for DFD, even though they never see a single DFD label. Building on this, the framework (Figure~\ref{fig:framework}) first adapts and then freezes the encoder. In Phase I Emotion-Guided Representation Alignment  (EmoBridge), pre-trained speech encoders are fine-tuned with an emotion recognition objective, guiding their representations toward affective cues while preserving their original semantic structure. The adapted encoders are then frozen in Phase II deepfake detection, where their outputs feed a lightweight classifier for DFD, so that no DFD gradient ever flows back to distort the learned representations.

\subsection{Phase I: Emotion-Guided Representation Alignment (EmoBridge)}
Let $f_\theta$ denote a pre-trained speech encoder with parameters $\theta$ initialised from $\theta^*_{pre}$, obtained by optimising an upstream task such as ASR or speaker verification on data $\mathcal{D}_{pre}$. In Phase I, we attach a lightweight emotion prediction head $g_\phi$(a SVM or a MLP) on top of $f_\theta$ and jointly fine-tune both on an emotion recognition task:
\begin{equation}
\label{eq:emo_bridge}
\theta^{*}_{\text{emo}},\, \phi^{*}
= \arg\min_{\theta,\,\phi}\;
\mathbb{E}_{(x,y)\sim\mathcal{D}_{\text{emo}}}
\big[\mathcal{L}_{\text{CE}}(g_{\phi}(f_{\theta}(x)),\, y)\big],
\;\; \text{s.t.}\; \theta \leftarrow \theta^{*}_{\text{pre}}
\end{equation}
where $\mathcal{L}_{\text{CE}}$ is the cross-entropy loss and $y$ is the emotion label. Crucially, the encoder is optimised \emph{solely} under the emotion objective: no deepfake label is ever observed in this phase. This is what distinguishes EmoBridge from conventional DFD fine-tuning—rather than steering the representation directly toward the DFD decision boundary, we steer it toward affective structure, which we show transfers to DFD as an emergent property.

Since our framework aims to be feature-agnostic and applies to any speech representation. We group commonly used features into three categories: hand-crafted acoustic features (e.g.\, those extracted by openSMILE~\cite{eyben2010opensmile}), deep learning (DL) raw features derived directly from neural encoders, and application-based features such as the semantic features of ASR models or the perceptual features of speaker verification (SV) models. For attention-based encoders we take the last hidden states, and for other architectures the final-layer outputs, as the representation of each modality. 

To demonstrate generality, we instantiate the framework with four representative features: hand-crafted acoustic features from openSMILE~\cite{eyben2010opensmile}; low-level DL features from Whisper\footnote{\url{https://huggingface.co/openai/whisper-small}}~\cite{radford2023robust}; semantic features from the ASR model SpeechT5\footnote{\url{https://huggingface.co/microsoft/speecht5\_asr}}~\cite{ao2022speecht5unifiedmodalencoderdecoderpretraining} and perceptual features from the SV model WavLM\footnote{\url{https://huggingface.co/microsoft/wavlm-base-sv}}~\cite{Chen_2022}. The deep encoders are taken directly from publicly available pre-trained checkpoints, which instantiate $\theta^*_{pre}$ in Eq.~\ref{eq:emo_bridge} rather than being trained from scratch; when no suitable pre-trained model exists for a target feature, an encoder can first be pre-trained and then incorporated into the same pipeline. These models are widely adopted in their respective domains, and we keep comparable configurations across experiments to ensure a fair comparison.

The two feature types require different treatment in Phase I. Hand-crafted features have no trainable encoder, so emotion guidance is realised through feature selection rather than fine-tuning. We compare two functional-level openSMILE feature sets: eGeMAPSv02, an expert-designed minimalistic set, and IS09, a challenge set that emphasises emotional cues beyond basic acoustics. Retaining the set richer in affective information lets the hand-crafted pathway approximate the same emotion-guided objective without parameter updates.

For deep encoders, we attach an emotion prediction head $g_\phi$, implemented as an MLP, on top of the encoder $f_\theta$, and jointly optimise both under the emotion objective in Eq.~\ref{eq:emo_bridge}. During this stage the encoder is updated only by the emotion loss, with no retention of its original pre-training objective and no exposure to DFD labels, so the representation is steered toward affective structure while its pre-trained semantic backbone is preserved.

\subsection{Phase~II: Frozen Representation for DFD}

Once Phase~I converges, the emotion prediction head $g_\phi$ has served its purpose and is discarded: it is merely the vehicle that drives the encoder toward affective structure during adaptation, and plays no role in detection. What we retain is solely the adapted encoder $f_{\theta^*_{emo}}$, which we freeze and use purely as a feature extractor. A lightweight classifier $h_{\varsigma}$ is trained for the real/fake decision:

  \begin{equation}
\label{eq:df}
\resizebox{0.92\columnwidth}{!}{$
\varsigma^{*}
= \arg\min_{\varsigma}\;
\mathbb{E}_{(x,y)\sim\mathcal{D}_{\text{df}}}
\big[\mathcal{L}_{\text{DFD}}(h_{\varsigma}(f_{\theta^{*}_{\text{emo}}}(x)),\, y)\big],
\;\; \text{s.t.}\; \theta^{*}_{\text{emo}}\;\text{frozen}
$}
\end{equation}

Discarding $g_\phi$ is conceptually important: the affective information that benefits DFD has already been internalised into the encoder's representation during PhaseI, rather than residing in the emotion classifier. The DFD gain therefore comes from the geometry of the adapted feature space itself, not from any explicit emotion prediction at test time. Freezing the encoder further guarantees that the DFD signal cannot back-propagate into the representation, so the measured performance reflects solely the quality of the emotion-aligned features. We deliberately keep $h_{\varsigma}$ lightweight, using a SVM\cite{cortes1995svm}, so that DFD accuracy is attributable to the representation rather than to classifier capacity. To probe where affective cues are most discriminative, we extract the output of each encoder layer and train a separate SVM per layer, reporting the average performance across all layers as a single, architecture-agnostic measure of how well the adapted representation supports DFD.

\section{Experiments and result}
\label{sec:guidelines}

\subsection{Dataset and experiments setup}

For Phase I, the emotion prediction head $g_\phi$is a three-layer fully connected network with hidden sizes of 768 and 256 followed by a 7-way output, optimised with AdamW at a learning rate of $10^{-5}$ for at most 40 epochs. In addition to adapting the ASR, SV, and DL-raw encoders, we include an encoder that is already emotion-aware, HuBERT\cite{hsu2021hubertselfsupervisedspeechrepresentation} pre-trained on IEMOCAP~\cite{busso2008iemocap}, as the \emph{Emotion} feature. This provides a fair reference point: it lets us test whether the DFD gain stems from the EmoBridge adaptation itself or simply from exposure to emotion, since this encoder natively encodes affect rather than acquiring it through Phase~I.

To learn emotion representations that generalise across speakers, recording conditions, and elicitation styles, we combine four widely used emotion recognition corpora for Phase I: TESS\cite{dupuis_pichora-fuller_2010_tess}, SAVEE~\cite{jackson_haq_savee}, CREMA-D~\cite{cao2014crema-d}, and RAVDESS~\cite{livingstone2018ravdess}. Training on their union, rather than any single corpus, reduces the risk that the adapted encoder overfits to dataset-specific acoustic or speaker characteristics. To verify that Phase I learns transferable affective structure rather than memorising the training corpora, we evaluate the learned representations on a held-out dataset, the Emotion Speech Dataset (ESD)\cite{zhou2021esd}, which is never seen during adaptation. Table~\ref{tab:emobridge} summarises the results. The adapted models reach competitive emotion recognition accuracy, confirming that Phase~I successfully steers the representations toward affect. They do not match state-of-the-art emotion recognisers, which is expected and in fact desirable: the encoders retain their original representational capacity rather than overfitting to emotion recognition.

\begin{table}[h]
\centering
\caption{EmoBridge performance. The table reports the weighted accuracy for the four pre-trained models evaluated in this study, with comparison to other models in literature.}
\label{tab:emobridge}
\begin{tabular}{lccccccc}
\toprule
Metrics & ASR & SV & DL-raw & Emotion & I. Shahin \cite{shahin2019emotion} & S. Hamsa \cite{hamsa2020emotion} & Koya S. \cite{koya2022ea} \\
\midrule
Weighted Acc. & 0.728 & 0.701 & 0.786 & 0.732 & 0.840 & 0.909 & 0.917 \\
\bottomrule
\end{tabular}%
\end{table}

After the EmoBridge step, Phase II evaluates DFD on three benchmarks: the ASVspoof2019 LA subset~\cite{wang2020asvspoof}, FoR~\cite{reimao2019dataset}, and ITW~\cite{muller2022does}. For ASVspoof~2019 LA and ITW we follow the official splits from the original papers, and for FoR we use the for-norm split, yielding approximately 71k, 32k, and 4.5k test samples, respectively. On each frozen encoder we train an SVM with the training data split and the default scikit-learn settings\footnote{\url{https://scikit-learn.org/stable/api/sklearn.svm.html}} on every layer, so that both low-level and high-level representations are probed; averaging across layers gives a single, architecture-agnostic measure of how well the adapted representation supports DFD. Performance is reported using accuracy and EER.

For comparison with the conventional DFD training strategy, we apply the original models directly to the corresponding training datasets and report the results on these datasets as in-domain performance. We additionally evaluate the same models on WaveFake \cite{frank2021wavefake} as an out-of-domain setting, following the same evaluation protocol described below.

\subsection{Results}
Table~\ref{tab:res} reports the layer-wise averaged EER and accuracy of the four models and acoustic features, evaluated on the FoR, ITW, and ASVSpoof2019 LA datasets. This evaluation protocol reflects the robustness of the representation across different abstraction levels and avoids bias to particular layers.

\begin{table}[ht]
\centering
\caption{Comparison of layer-wise averaged EER and accuracy between the pre-trained model, EmoBridge, WaveFake fine-tuned encoders, and in-domain fine-tuned encoders on three datasets.}
\label{tab:res}
\resizebox{0.8\columnwidth}{!}{
\begin{tabular}{p{2cm}c|cc|cc|cc|cc}
\toprule
\multirow{2}{*}{\textbf{Dataset}} & 
\multirow{2}{*}{\textbf{Model}} &
\multicolumn{2}{c|}{\textbf{Pre-trained}} &
\multicolumn{2}{c|}{\textbf{EmoBridge}} &
\multicolumn{2}{c|}{\textbf{WaveFake FT}} &
\multicolumn{2}{c}{\textbf{In-domain FT}} \\
\cmidrule(lr){3-4}
\cmidrule(lr){5-6}
\cmidrule(lr){7-8}
\cmidrule(lr){9-10}
& & \textbf{EER$\downarrow$} & \textbf{Acc$\uparrow$}
& \textbf{EER$\downarrow$} & \textbf{Acc$\uparrow$}
& \textbf{EER$\downarrow$} & \textbf{Acc$\uparrow$}
& \textbf{EER$\downarrow$} & \textbf{Acc$\uparrow$} \\
\midrule

\multirow{5}{*}{\textbf{FoR}}
& openSMILE 
& .406 & .574 & \textbf{.275} & \textbf{.663} 
& -- & -- & -- & -- \\
& Emotion 
& .139 & .854 & \textbf{.078} & \textbf{.913} 
& .110 & .875 & .211 & .632 \\
& SV      
& .136 & .858 & \textbf{.089} & \textbf{.908} 
& .126 & .861 & .144 & .719 \\
& ASR     
& .082 & .887 & \textbf{.045} & \textbf{.946} 
& .091 & .855 & .218 & .702 \\
& DL-raw  
& .091 & .890 & \textbf{.062} & \textbf{.931} 
& .089 & .898 & .056 & .886 \\

\midrule

\multirow{5}{*}{\textbf{ITW}}
& openSMILE 
& .200 & .897 & \textbf{.188} & .897 
& -- & -- & -- & -- \\
& Emotion 
& .050 & .953 & .047 & .953 
& .032 & .971 & \textbf{.003} & \textbf{.997} \\
& SV      
& .061 & .942 & .058 & .946 
& .040 & .962 & \textbf{.004} & \textbf{.997} \\
& ASR     
& .054 & .948 & .052 & .947 
& .040 & .962 & \textbf{.003} & \textbf{.998} \\
& DL-raw  
& .033 & .969 & .020 & .982 
& .030 & .974 & \textbf{.012} & \textbf{.991} \\

\midrule

\multirow{5}{*}{\parbox{1.2cm}{\centering\textbf{ASVspoof\\2019 LA}}}
& openSMILE 
& .197 & .896 & \textbf{.188} & \textbf{.898} 
& -- & -- & -- & -- \\
& Emotion 
& .061 & .970 & .071 & .966 
& .035 & .985 & \textbf{.022}&\textbf{.983} \\
& SV      
& .070 & .964 & .079 & .960 
& .045 & .976 & \textbf{.023} & \textbf{.987} \\
& ASR     
& .076 & .960 & .078 & .960 
& .051 & .969 & \textbf{.024} &\textbf{.983} \\
& DL-raw  
& .052 & .946 & .049 & .967 
& .074 & .916 & \textbf{.024} & \textbf{.974} \\

\bottomrule
\end{tabular}}
\end{table}

This demonstrates that EmoBridge consistently improves deepfake detection over the pre-trained baseline across most datasets and feature types, achieving lower EER and higher accuracy by providing complementary emotion cues. Compared with the emotion-pretrained HuBERT model, the pre-EmoBridge performance is not consistently superior, likely due to the limited emotional diversity in IEMOCAP. By incorporating broader and more balanced emotion knowledge, EmoBridge further enhances discrimination. Although EmoBridge does not surpass fully supervised DFD fine-tuning, it achieves competitive performance without task-specific encoder optimisation, suggesting that it improves deepfake discrimination while preserving general speech representations. Such preservation is crucial for maintaining broader speech capabilities, which will be analysed in subsequent experiments. Across datasets, DL-raw features consistently benefit from EmoBridge, confirming that deep representations capture effective DFD cues and can be enhanced by emotion guidance. FoR shows the largest gains, with substantial EER reductions and accuracy improvements, while ITW also benefits despite smaller relative improvements due to its large scale. These results indicate that emotion-based bridging is particularly effective under diverse and expressive speech conditions.

In contrast, comparable performance is observed on ASVSpoof. To identify where our approach is most vulnerable, we further analyse model performance across different sources within the test set. We also compare against two existing emotion-based methods: using the static emotion feature \cite{conti2022deepfake} (results reported directly from the original paper), and a concatenating emotion feature with others (adapted from \cite{mittal2020emotions}), which we combine HuBERT embeddings with DL-raw features (using our same experimental configuration of the models in Section 3.1). The results are shown in Table~\ref{tab:asvspoof}.

\begin{table}[ht]
\centering
\caption{Accuracy comparison across different sources and feature settings before and after the emotion as a bridge strategy. Column Static is from work \cite{mittal2020emotions} (--- means not provided or N/A) while Column Emotion is from our strategy using Hubert as original feature. In this table, only the most noteworthy results are emphasised in bold.}
\label{tab:asvspoof}
\resizebox{0.8\columnwidth}{!}{%
\begin{tabular}{lcccccccccc}
\toprule
\textbf{Source} & \multicolumn{2}{c}{\textbf{ASR}} & \multicolumn{2}{c}{\textbf{SV}} & \multicolumn{2}{c}{\textbf{DL-raw}} & \multicolumn{2}{c}{\textbf{Emotion}} & \textbf{Static} & \textbf{Concatenation}  \\
\cmidrule(lr){2-3} \cmidrule(lr){4-5} \cmidrule(lr){6-7} \cmidrule(lr){8-9} \cmidrule(l){10-10} \cmidrule(l){11-11}
\textbf{EmoBridge?} & \textbf{No} & \textbf{Yes} & \textbf{No} & \textbf{Yes} & \textbf{No} & \textbf{Yes} & \textbf{No} & \textbf{Yes} & \textbf{---}  & \textbf{---} \\
\midrule
A07  & 1.000  & 0.999  & 0.999  & 1.000  & 0.998  & 0.999  & 1.000  & 0.998 & 0.948 & 0.994 \\
A08  & 0.995  & \textbf{1.000}  & 0.999  & 0.999  & 0.949  & \textbf{0.995}  & 0.988  & \textbf{0.999} & 0.988 & 0.962 \\
A09  & 1.000  & 1.000  & 1.000  & 1.000  & 1.000  & 1.000  & 1.000  & 1.000 & 1.000 & 1.000 \\
A10  & 0.999  & 0.996  & 0.995  & \textbf{0.996}  & 0.842  & \textbf{0.857}  & 0.998  & 0.977 & 0.900 & 0.965 \\
A11  & 1.000  & 1.000  & 1.000  & 1.000  & 0.979  & 0.937  & 1.000  & 0.999 & 0.895 & 0.997 \\
A12  & 1.000  & 0.998  & 1.000  & 1.000  & 0.999  & 0.991  & 1.000  & 0.998 & 0.890 & 1.000 \\
A13  & 1.000  & 1.000  & 1.000  & 1.000  & 1.000  & 1.000  & 1.000  & 1.000 & 0.831 & 1.000 \\
A14  & 1.000  & 1.000  & 1.000  & 1.000  & 0.987  & 0.990  & 1.000  & 1.000 & 0.763 & 0.952 \\
A15  & 1.000  & 0.986  & 0.998  & 0.998  & 0.996  & 0.998  & 1.000  & 0.999 & 0.927 & 0.954 \\
A16  & 0.999  & 0.995  & 0.997  & 0.996  & 0.994  & 0.994  & 0.999  & 0.993 & 0.898 & 0.988 \\
A17  & 0.898  & \textbf{0.980}  & 0.968  & 0.956  & 0.999  & 0.999  & 0.954  & \textbf{0.984} & --- & 0.988 \\
A18  & 0.956  & \textbf{0.996}  & 0.994  & 0.990  & 0.948  & \textbf{0.969}  & 0.793  & \textbf{0.975} & --- & 0.869 \\
A19  & 0.788  & \textbf{0.920}  & 0.780  & \textbf{0.890}  & 0.966  & 0.941  & 0.813  & \textbf{0.903} & --- & 0.905 \\
Real & 0.695  & 0.520  & 0.749  & 0.486  & 0.860  & \textbf{0.884}  & 0.820  & 0.564 & --- & 0.832 \\
\bottomrule
\end{tabular}

}
\end{table}

On ASVspoof2019 LA, the gains are smaller than on FoR and ITW. A closer look shows this is not merely an effect of the imbalanced bona fide/spoof distribution, but a \emph{speaker-dependent} bias in the bona fide subset. Per-speaker error rates after EmoBridge adaptation span a wide range, from a 10.3\% floor to 89.0\% (Table \ref{tab:speaker_bias}): the distribution is heavy-tailed, with 26\% of speakers accounting for nearly half of all misclassifications. Such concentration indicates a speaker-specific artefact rather than a uniform degradation of the representation, since a genuine flaw of the method would affect all speakers comparably. Consistently, this pattern is specific to ASVspoof and does not appear on FoR or ITW, where real speech remains accurately detected.

\begin{table}[h]
\centering
\caption{Distribution of per-speaker bona fide error rates on ASVspoof~2019 LA after EmoBridge adaptation (Emotion feature, layer-averaged). Degradation is heavy-tailed: 26\% of speakers account for 49\% of all misclassifications, while even the least-affected speakers retain a $\sim$10\% error floor.}
\label{tab:speaker_bias}
\resizebox{0.8\columnwidth}{!}{%
\begin{tabular}{lcccc}
\toprule
\textbf{Error band} & \textbf{\#Spk.} & \textbf{\#Samples} & \textbf{\#Misclass.} & \textbf{Rate} \\
\midrule
$\geq$50\% (severe)   & 17 & 1{,}963 & 1{,}285 & 65.5\% \\
30--50\% (moderate)   & 13 & 1{,}556 & 589     & 37.9\% \\
10--30\% (mild)       & 35 & 3{,}643 & 725     & 19.9\% \\
\midrule
\textbf{Overall}      & \textbf{65} & \textbf{7{,}162} & \textbf{2{,}599} & \textbf{36.3\%} \\
\bottomrule
\end{tabular}%

}
\end{table}
\vspace{-1cm}
\subsection{Ruling Out Alternative Bridges}

The degradation is concentrated on specific speakers, which raises a concern: perhaps other features such as speaker characteristics, could also serve as the effective bridge for DFD. To test this, we run a control experiment that replaces the emotion objective while keeping everything else fixed. This control, \textbf{SpeakerBridge}, fine-tunes the encoder to classify speaker identity, a more fine-grained but non-affective task. We single out speaker identity because prior work has established a strong coupling between emotional and speaker-related cues in speech\cite{schuller2013computational}, making it the most competitive non-affective candidate for explaining the gain. 

As shown in Table~\ref{tab:control}, SpeakerBridge \emph{degrades} DFD performance relative to the pre-trained baseline. This confirms that the benefit is specific to the emotion objective and does not arise merely from the greater task complexity of speaker classification. Rather than exhaustively evaluating every conceivable auxiliary task, we focus on speaker identity as arguably the most informative control. It is among the tasks most strongly entangled with emotion in the literature, while few other tasks appear to share both emotion's close coupling with the original speech representation and its established relevance to deepfake detection. Emotion thus acts as a natural bridge between the original speech representation and DFD that cannot be substituted by related tasks.

\begin{table}[ht]
\centering
\vspace{-0.5cm}
\caption{Comparison of layer-wise averaged EER and accuracy between the Pre-trained model and the model with the speaker as a bridge strategy on three datasets.}
\label{tab:control}
\begin{tabular}{p{2cm}cccccc}
\toprule
\multirow{2}{*}{\textbf{Dataset}} & \multirow{2}{*}{\textbf{Model}}  & \multicolumn{2}{c}{\textbf{Pre-trained}} & \multicolumn{2}{c}{\textbf{SpeakerBridge}} \\
\cmidrule(lr){3-4} \cmidrule(lr){5-6}
 &  & \textbf{EER $\downarrow$} & \textbf{Acc $\uparrow$} & \textbf{EER $\downarrow$} & \textbf{Acc $\uparrow$} \\
\midrule
\multirow{4}{*}{\textbf{FoR}} 
& Emotion & \textbf{.139} & .854 & .148 & \textbf{.857} \\
& SV      & \textbf{.136} & \textbf{.858} & .136 & .854 \\
& ASR     & \textbf{.082} & \textbf{.887} & .135 & .884 \\
& DL-raw  & \textbf{.091} & \textbf{.890} & .137 & .775 \\
\midrule
\multirow{4}{*}{\textbf{ITW}} 
& Emotion & \textbf{.050} & \textbf{.953} & .148 & .870 \\
& SV      & \textbf{.061} & \textbf{.942} & .140 & .876 \\
& ASR     & \textbf{.054} & \textbf{.948} & .139 & .876 \\
& DL-raw  & .033 & .969 & \textbf{.020} & \textbf{.983} \\
\midrule
\multirow{4}{*}{\parbox{1cm}{\centering\textbf{ASVSpoof\\ 2019 LA}}}
& Emotion & \textbf{.061} & \textbf{.970} & .155 & .864 \\
& SV      & \textbf{.070} & \textbf{.964} & .151 & .872 \\
& ASR     & \textbf{.076} & \textbf{.960} & .150 & .873 \\
& DL-raw  & \textbf{.052} & .946 & .063 & \textbf{.964} \\
\bottomrule
\end{tabular}
\vspace{-1.2cm}
\end{table}

\subsection{Effects of DFD classifier}

We select SVM as the main classifier for simplicity, although other classifiers could also be used. Previous work shows that the relative performance of different feature representations remains consistent across classifier choices~\cite{chakravarty2024lightweight}, even though absolute scores may vary. To further validate the robustness of our approach, we conduct an ablation study on the FoR dataset using a two-layer MLP classifier (hidden dimension 512 with a 2-dimensional output), as shown in Table~\ref{tab:abl_mlp}. The results show similar performance across feature types, demonstrating the robustness of our method.

\begin{table}[ht]
\centering
\caption{EER and accuracy between the Pre-trained model and the model with the EmoBridge strategy on FoR with MLP as classifier.}
\label{tab:abl_mlp}
\begin{tabular}{p{1cm}cccccc}
\toprule
\multirow{2}{*}{\textbf{Dataset}} & \multirow{2}{*}{\textbf{Model}}  & \multicolumn{2}{c}{\textbf{Pre-trained}} & \multicolumn{2}{c}{\textbf{EmoBridge}} \\
\cmidrule(lr){3-4} \cmidrule(lr){5-6}
 &  & \textbf{EER $\downarrow$} & \textbf{Acc $\uparrow$} & \textbf{EER $\downarrow$} & \textbf{Acc $\uparrow$} \\
\midrule
\multirow{5}{*}{\textbf{FoR}} 
& openSMILE & .385 & .554 & \textbf{.265} & \textbf{.645} \\
& Emotion & .143 & .866 & \textbf{.083} & \textbf{.915} \\
& SV      & .146 & .864 & \textbf{.087} & \textbf{.910} \\
& ASR     & .088 & .898 & \textbf{.066} & \textbf{.952} \\
& DL-raw  & .095 & .899 & \textbf{.077} & \textbf{.943} \\

\bottomrule
\end{tabular}

\vspace{-0.5cm}
\end{table}
\subsection{Preservation of speech representation}
\begin{figure}[t]
    \centering
    \includegraphics[width=0.75\linewidth]{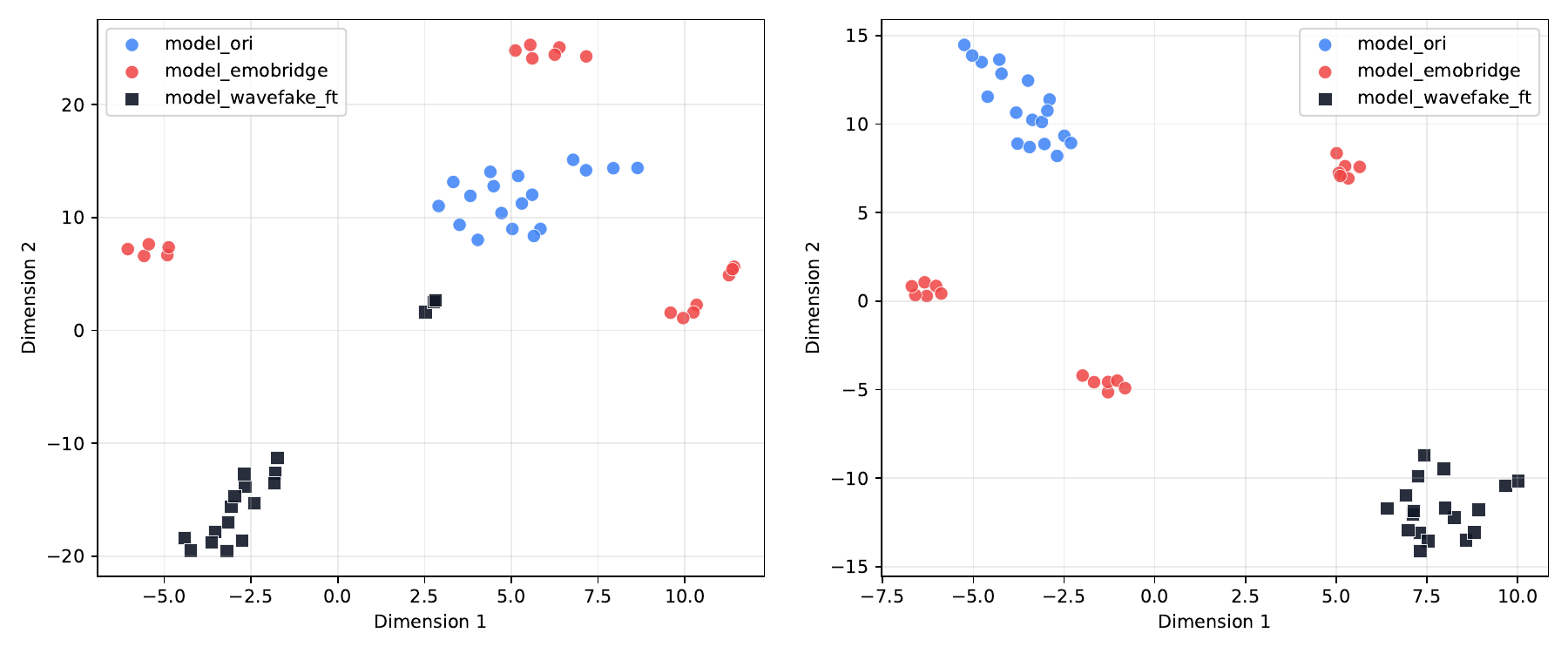}
    \caption{t-SNE visualization of ASR (left) and SV (right) representations from selected samples obtained with the original model (\textit{model\_ori}), after EmoBridge training (\textit{model\_emobridge}), and after conventional DFD fine-tuning (model\_wavefake\_ft).}
    \label{fig:sv-tsne-comparison}
    \vspace{-0.5cm}
\end{figure}

Additionally, we evaluated the preservation of speech representation via feature visualization using selected samples from the EmoFake~\cite{zhao2024emofake}. These samples are controlled to share the same speaker, content, or emotion, allowing factor-wise comparison of feature representations. We analyze SV and ASR features, as shown in Figure~\ref{fig:sv-tsne-comparison},  using t-SNE~\cite{maaten2008visualizing} for visualization.

After applying EmoBridge, the features exhibit clearer emotion-aware clustering, demonstrating the effectiveness of emotion guidance. Meanwhile, the overall feature distribution remains highly consistent with the original representations, as indicated by the close alignment between \textit{model\_emobridge} and \textit{model\_ori}. In contrast, conventional DFD fine-tuning causes a huge shift in the representation space, suggesting that direct optimisation for deepfake discrimination may alter the underlying speech representations.


We additionally assess representation preservation through encoder-level, task-agnostic evaluation. Specifically, we evaluate both SV and ASR performance in a zero-shot setting, comparing the original models with their counterparts trained using EmoBridge and conventional DFD training on WaveFake. For SV, we evaluate the SV model on VoxCeleb~\cite{nagrani2020voxceleb}, with EERs reported in Table~\ref{tab:representation_preservation}; for ASR, we evaluate ASR model on LibriSpeech \cite{panayotov2015librispeech}, with Word error rate (WER) reported in Table~\ref{tab:representation_preservation}. These results indicate that EmoBridge effectively improves deepfake detection while largely preserving the general speech representations learned during pre-training. In contrast, conventional DFD fine-tuning causes substantial representation drift and severely degrades both speaker verification and speech recognition performance. For example, the >100\% WER reflects severe insertion errors as the encoder representation collapses under DFD supervision, verified by manual inspection of a random 50-sample subset. This demonstrates that EmoBridge avoids over-specialising the encoder towards a single detection objective, providing scalability and keeping speech capabilities.

\begin{table}[h]
\centering
\caption{Representation preservation evaluated on speaker verification (SV) and automatic speech recognition (ASR). Lower EER and WER indicate better preservation.}
\label{tab:representation_preservation}
\begin{tabular}{lllc}
\toprule
Task & Model & Training Strategy & Metric $\downarrow$ \\
\midrule
\multirow{3}{*}{SV}
& \multirow{3}{*}{WavLM-base-SV}
& Pretrained & 5.63\% \\
& & EmoBridge fine-tuned & 6.78\% \\
& & Conventional DFD & 35.13\% \\
\midrule
\multirow{3}{*}{ASR}
& \multirow{3}{*}{SpeechT5-ASR}
& Pretrained & 11.91\% \\
& & EmoBridge fine-tuned & 16.56\% \\
& & Conventional DFD & 221.4\% \\
\bottomrule
\end{tabular}
\end{table}

\subsection{Explainability for model improvement}

\begin{figure}[h]
\vspace{-0.3cm}
    \centering
    \includegraphics[width=0.6\linewidth]{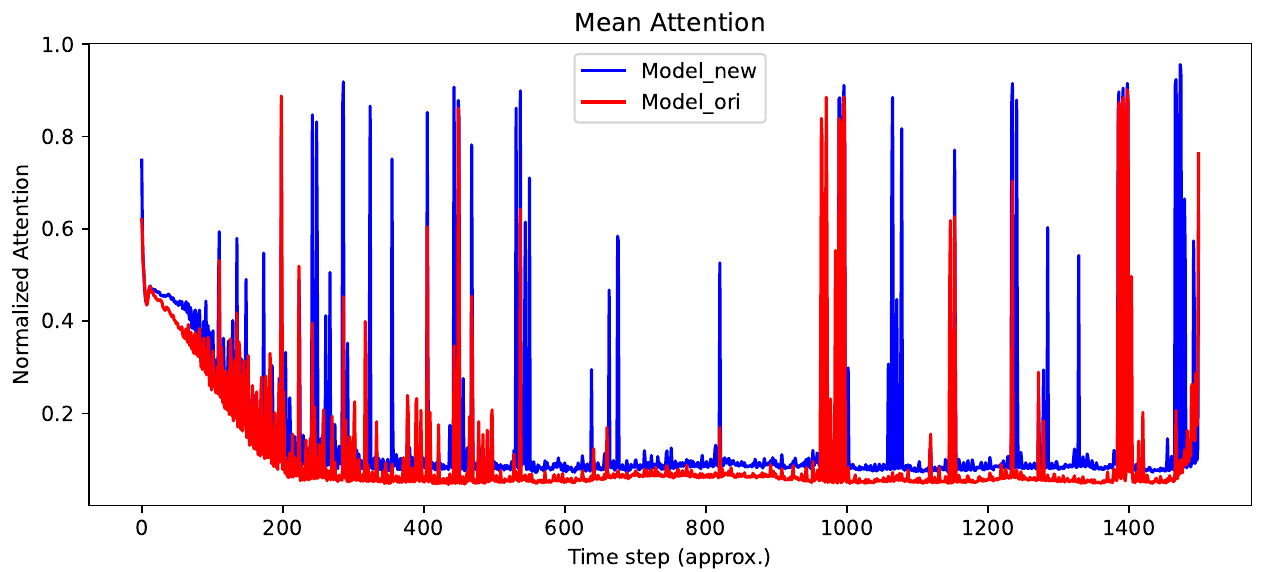}
    \caption{Comparison of mean attention values of Whisper layers before (model\_ori) and after emotion as a bridge (model\_new).}
    \label{fig:raw-attn}
    \vspace{-0.6cm}
\end{figure}

Our approach yields performance gains, so we further investigate the model’s internal mechanism. Specifically, we examine the mean attention values across time steps for different layers before and after applying our strategy, using the DL-raw features as an example, shown in Figure \ref{fig:raw-attn}. The results show that the model after applying our strategy retains partial overlap with the original attention distribution, indicating that it preserves useful prior information, while also developing new attention regions where more affective cues are captured.

\section{Conclusion}

We propose EmoBridge, a unified strategy that leverages emotion information as a constraint to guide pre-trained feature embeddings. This approach preserves the original feature representations while injecting affective cues, enabling the model to better expose discriminative patterns for DFD and consistently outperform baseline methods. Our experiments show that emotion information provides the effective guidance for improving representation quality. Future work could explore integrating multimodal features with emotion-aware guidance for DFD.
\newpage

{\footnotesize
\bibliographystyle{splncs04}
\bibliography{reference}

@article{hamza2022deepfake,
  title={Deepfake audio detection via {MFCC} features using machine learning},
  author={Hamza, Ameer and Javed, Abdul Rehman and Iqbal, Farkhund and others},
  journal={IEEE Access},
  volume={10},
  pages={134018--134028},
  year={2022}
}

@book{schuller2013computational,
  title={Computational Paralinguistics: Emotion, Affect and Personality in Speech and Language Processing},
  author={Schuller, Bj{\"o}rn and Batliner, Anton},
  year={2013},
  publisher={Wiley}
}

@inproceedings{bakken2025deep,
  title={Deep Fake Audio Detection Framework Using {MFCC}s, Chroma Features, and Spectrogram Images},
  author={Bakken, Nora and Singh, Shavil and Prashant, Makwana and others},
  booktitle={IEEE Conf. on Artificial Intelligence (CAI)},
  year={2025}
}

@inproceedings{frank2021wavefake,
  title={{WaveFake}: A Data Set to Facilitate Audio Deepfake Detection},
  author={Frank, Joel and Sch{\"o}nherr, Lea},
  booktitle={NeurIPS},
  year={2021}
}

@inproceedings{panayotov2015librispeech,
  title={{Librispeech}: An {ASR} Corpus Based on Public Domain Audio Books},
  author={Panayotov, Vassil and Chen, Guoguo and Povey, Daniel and others},
  booktitle={ICASSP},
  pages={5206--5210},
  year={2015}
}

@inproceedings{eyben2010opensmile,
  title={{openSMILE}: the {M}unich versatile and fast open-source audio feature extractor},
  author={Eyben, Florian and W{\"o}llmer, Martin and Schuller, Bj{\"o}rn},
  booktitle={ACM Multimedia},
  pages={1459--1462},
  year={2010}
}

@inproceedings{radford2023robust,
  title={Robust speech recognition via large-scale weak supervision},
  author={Radford, Alec and Kim, Jong Wook and Xu, Tao and others},
  booktitle={ICML},
  year={2023}
}

@inproceedings{mansoor2024audio,
  title={Audio Deepfake Detection: End-to-End training with powerful pretrained {ASR}},
  author={Mansoor, Kh Hamad and Alam, Mehreen},
  booktitle={INMIC},
  year={2024}
}

@inproceedings{pianese2022deepfake,
  title={Deepfake audio detection by speaker verification},
  author={Pianese, Alessandro and Cozzolino, Davide and Poggi, Giovanni and others},
  booktitle={IEEE WIFS},
  year={2022}
}

@misc{sun2024friendlyaicomprehensivereview,
  title={Towards Friendly {AI}: A Comprehensive Review and New Perspectives on Human-{AI} Alignment},
  author={Sun, Qiyang and Li, Yupei and Alturki, Emran and others},
  year={2024},
  note={arXiv:2412.15114}
}

@inproceedings{conti2022deepfake,
  title={Deepfake speech detection through emotion recognition: a semantic approach},
  author={Conti, Emanuele and Salvi, Davide and Borrelli, Clara and others},
  booktitle={ICASSP},
  pages={8962--8966},
  year={2022}
}

@inproceedings{tits2018asr,
  title={{ASR}-based Features for Emotion Recognition: A Transfer Learning Approach},
  author={Tits, No{\'e} and El Haddad, Kevin and Dutoit, Thierry},
  booktitle={Challenge-HML Workshop, ACL},
  pages={48--52},
  year={2018}
}

@misc{tian2025marcovoicetechnicalreport,
  title={{Marco-Voice} Technical Report},
  author={Tian, Fengping and Lyu, Chenyang and Ni, Xuanfan and others},
  year={2025},
  note={arXiv:2508.02038}
}

@inproceedings{mittal2020emotions,
  title={Emotions don't lie: An audio-visual deepfake detection method using affective cues},
  author={Mittal, Trisha and Bhattacharya, Uttaran and Chandra, Rohan and others},
  booktitle={ACM Multimedia},
  pages={2823--2832},
  year={2020}
}

@inproceedings{hosler2021deepfakes,
  title={Do deepfakes feel emotions? A semantic approach to detecting deepfakes via emotional inconsistencies},
  author={Hosler, Brian and Salvi, Davide and Murray, Anthony and others},
  booktitle={CVPR Workshops},
  pages={1013--1022},
  year={2021}
}

@inproceedings{reimao2019dataset,
  title={{FoR}: A dataset for synthetic speech detection},
  author={Reimao, Ricardo and Tzerpos, Vassilios},
  booktitle={SpeD},
  year={2019}
}

@article{muller2022does,
  title={Does audio deepfake detection generalize?},
  author={M{\"u}ller, Nicolas M. and Czempin, Pavel and Dieckmann, Franziska and others},
  journal={Interspeech},
  year={2022}
}

@article{wang2020asvspoof,
  title={{ASVspoof} 2019: A large-scale public database of synthesized, converted and replayed speech},
  author={Wang, Xin and Yamagishi, Junichi and Todisco, Massimiliano and others},
  journal={Comput. Speech Lang.},
  volume={64},
  pages={101114},
  year={2020}
}

@article{alsaeedi2025audio,
  title={Audio-Visual Multimodal Deepfake Detection Leveraging Emotional Recognition},
  author={Alsaeedi, Alaa and AlMansour, Amal and Jamal, Amani},
  journal={Int. J. Adv. Comput. Sci. Appl.},
  volume={16},
  number={6},
  year={2025}
}

@inproceedings{ao2022speecht5unifiedmodalencoderdecoderpretraining,
  title={{SpeechT5}: Unified-modal encoder-decoder pre-training for spoken language processing},
  author={Ao, Junyi and Wang, Rui and Zhou, Long and others},
  booktitle={ACL},
  pages={5723--5738},
  year={2022}
}

@article{Chen_2022,
  title={{WavLM}: Large-Scale Self-Supervised Pre-Training for Full Stack Speech Processing},
  author={Chen, Sanyuan and Wang, Chengyi and Chen, Zhengyang and others},
  journal={IEEE J. Sel. Top. Signal Process.},
  volume={16},
  number={6},
  pages={1505--1518},
  year={2022}
}

@misc{dupuis_pichora-fuller_2010_tess,
  title={Toronto Emotional Speech Set ({TESS})},
  author={Dupuis, Kate and Pichora-Fuller, M. Kathleen},
  year={2010},
  note={Univ. Toronto dataset}
}

@misc{jackson_haq_savee,
  title={Surrey Audio-Visual Expressed Emotion ({SAVEE}) Database},
  author={Jackson, Philip and Haq, Sanaul},
  note={Univ. Surrey dataset}
}

@article{cao2014crema-d,
  title={{CREMA-D}: Crowd-sourced Emotional Multimodal Actors Dataset},
  author={Cao, Houwei and Cooper, David G. and Keutmann, Michael K. and others},
  journal={IEEE Trans. Affect. Comput.},
  volume={5},
  number={4},
  pages={377--390},
  year={2014}
}

@article{livingstone2018ravdess,
  title={The Ryerson Audio-Visual Database of Emotional Speech and Song ({RAVDESS})},
  author={Livingstone, Steven R. and Russo, Frank A.},
  journal={PLOS ONE},
  volume={13},
  number={5},
  pages={e0196391},
  year={2018}
}

@inproceedings{zhou2021esd,
  title={Seen and unseen emotional style transfer for voice conversion with a new emotional speech dataset},
  author={Zhou, Kun and Sisman, Berrak and Liu, Rui and others},
  booktitle={ICASSP},
  pages={920--924},
  year={2021}
}

@article{cortes1995svm,
  title={Support-Vector Networks},
  author={Cortes, Corinna and Vapnik, Vladimir},
  journal={Mach. Learn.},
  volume={20},
  number={3},
  pages={273--297},
  year={1995}
}

@article{hsu2021hubertselfsupervisedspeechrepresentation,
  title={{HuBERT}: Self-Supervised Speech Representation Learning by Masked Prediction of Hidden Units},
  author={Hsu, Wei-Ning and Bolte, Benjamin and Tsai, Yao-Hung Hubert and others},
  journal={IEEE/ACM Trans. Audio Speech Lang. Process.},
  volume={29},
  pages={3451--3460},
  year={2021}
}

@article{busso2008iemocap,
  title={{IEMOCAP}: Interactive emotional dyadic motion capture database},
  author={Busso, Carlos and Bulut, Murtaza and Lee, Chi-Chun and others},
  journal={Lang. Resour. Eval.},
  volume={42},
  number={4},
  pages={335--359},
  year={2008}
}

@inproceedings{zhao2024emofake,
  title={{EmoFake}: An initial dataset for emotion fake audio detection},
  author={Zhao, Yan and Yi, Jiangyan and Tao, Jianhua and others},
  booktitle={CCL},
  pages={419--433},
  year={2024}
}

@article{maaten2008visualizing,
  title={Visualizing data using {t-SNE}},
  author={van der Maaten, Laurens and Hinton, Geoffrey},
  journal={J. Mach. Learn. Res.},
  volume={9},
  pages={2579--2605},
  year={2008}
}

@article{shahin2019emotion,
  title={Emotion recognition using hybrid {G}aussian mixture model and deep neural network},
  author={Shahin, Ismail and Nassif, Ali Bou and Hamsa, Shibani},
  journal={IEEE Access},
  volume={7},
  pages={26777--26787},
  year={2019}
}

@article{hamsa2020emotion,
  title={Emotion recognition from speech using wavelet packet transform cochlear filter bank and random forest classifier},
  author={Hamsa, Shibani and Shahin, Ismail and Iraqi, Youssef and others},
  journal={IEEE Access},
  volume={8},
  pages={96994--97006},
  year={2020}
}

@inproceedings{koya2022ea,
  title={{EA-VGG}: A new approach for emotional speech classification},
  author={Koya, Shibani Hamsa and Shahin, Ismail and Iraqi, Youssef and others},
  booktitle={ICECCME},
  year={2022}
}

@misc{li2025artificialemotionsurveytheories,
  title={Artificial Emotion: A Survey of Theories and Debates on Realising Emotion in Artificial Intelligence},
  author={Li, Yupei and Sun, Qiyang and Schlicher, Michelle and others},
  year={2025},
  note={arXiv:2508.10286}
}

@inproceedings{luo2024whisper,
  title={{Whisper+AASIST} for DeepFake Audio Detection},
  author={Luo, Qian and Vinayagam Sivasundari, Kalyani},
  booktitle={HCI Int'l},
  pages={121--133},
  year={2024}
}

@article{xu2025qwen2,
  title={{Qwen2.5-Omni} technical report},
  author={Xu, Jin and Guo, Zhifang and He, Jinzheng and others},
  journal={arXiv:2503.20215},
  year={2025}
}

@article{nagrani2020voxceleb,
  title={{VoxCeleb}: Large-scale speaker verification in the wild},
  author={Nagrani, Arsha and Chung, Joon Son and Xie, Weidi and others},
  journal={Comput. Speech Lang.},
  volume={60},
  pages={101027},
  year={2020}
}

@article{chakravarty2024lightweight,
  title={A lightweight feature extraction technique for deepfake audio detection},
  author={Chakravarty, Nidhi and Dua, Mohit},
  journal={Multimed. Tools Appl.},
  volume={83},
  number={26},
  pages={67443--67467},
  year={2024}
}

@article{wang2025awaveformer,
  title={{AWaveFormer}: Audio Wavelet Transformer Network for Generalized Audio Deepfake Detection},
  author={Wang, Rui and Chen, Zirui and Wang, Bo and others},
  journal={IEEE Trans. Audio Speech Lang. Process.},
  year={2025}
}

@article{khanjani2023audio,
  title={Audio deepfakes: A survey},
  author={Khanjani, Zahra and Watson, Gabrielle and Janeja, Vandana P.},
  journal={Front. Big Data},
  volume={5},
  pages={1001063},
  year={2023}
}

@inproceedings{saha2024exploring,
  title={Exploring green {AI} for audio deepfake detection},
  author={Saha, Subhajit and Sahidullah, Md and Das, Swagatam},
  booktitle={EUSIPCO},
  pages={186--190},
  year={2024}
}

@misc{li2025gatedxlstmmultimodalaffectivecomputing,
  title={{GatedxLSTM}: A Multimodal Affective Computing Approach for Emotion Recognition in Conversations},
  author={Li, Yupei and Sun, Qiyang and Krishna Murthy, Sunil Munthumoduku and others},
  year={2025},
  note={arXiv:2503.20919}
}
}

\end{document}